\documentclass[rapids]{jfm}
\usepackage{natbib}

\usepackage[utf8]{inputenc}
\usepackage[T1]{fontenc}
\usepackage{xcolor, mathtools, graphicx, dcolumn, bm, tikz, pgfplots, afterpage, capt-of, pdflscape}
\usepackage[colorlinks=true,citecolor=blue,linkcolor=blue]{hyperref}

\usetikzlibrary{intersections}
\usetikzlibrary{patterns}

\newcommand{\eq}[2]{\begin{equation} \label{eq:#1} #2 \end{equation}}

\newcommand{\bs}[1]{\boldsymbol{#1}}
\def\be{\begin{equation}}
\def\ee{\end{equation}}
\def\ba{\begin{eqnarray}}
\def\ea{\end{eqnarray}}

\def\pdt{\partial_t}
\def\div{\bnabla \cdot}
\def\rot{\bnabla \times}
\def\dd{\mathrm{d}}

\begin{document}

\title{Locality of triad interaction and Kolmogorov constant in inertial wave turbulence}
\shorttitle{Locality and Kolmogorov constant in inertial wave turbulence}

\author{Vincent David \aff{1,2}
  \corresp{\email{vincent.david@lpp.polytechnique.fr}}
 \and Sébastien Galtier \aff{1,2,3}}

\affiliation{\aff{1}Laboratoire de Physique des Plasmas, \'Ecole polytechnique, F-91128 Palaiseau Cedex, France
\aff{2}Universit\'e Paris-Saclay, IPP, CNRS, Observatoire Paris-Meudon, France
\aff{3}Institut universitaire de France}

\maketitle

\begin{abstract}
Using the theory of wave turbulence for rapidly rotating incompressible fluids derived by Galtier (2003), we find the locality conditions that the solutions of the kinetic equation must satisfy. We show that the exact anisotropic Kolmogorov-Zakharov (KZ) spectrum satisfies these conditions, which justifies the existence of this constant (positive) energy flux solution. Although a direct cascade is predicted in the transverse ($\perp$) and parallel ($\parallel$) directions to the rotation axis, we show numerically that in the latter case some triadic interactions can have a negative contribution to the energy flux, while in the former case all interactions contribute to a positive flux. Neglecting the parallel energy flux, we estimate the Kolmogorov constant at $C_K \simeq 0.749$. These results provide theoretical support for recent numerical and experimental studies. 
\end{abstract}

\section{Introduction}\label{sec:Introduction}
Wave turbulence theory describes a set of random waves in weakly nonlinear interactions \citep{Zakharov_1992,Nazarenko_2011,Galtier_2023}. The strength of this theory, based on a multiple-scale technique, lies in its analytical and rigorous character, and the fact that it leads to a natural asymptotic closure of the hierarchy of moment equations when considering the long-term statistical behaviour. We are interested here in waves resulting from the rapid rotation, at a rate $\Omega$, of an incompressible fluid: these are inertial waves. They appear when the Coriolis force is introduced into the Navier-Stokes equations, which breaks the spherical symmetry and introduce (statistical) anisotropy.
The theory of inertial wave turbulence was derived by \cite{Galtier_2003a}. It is an asymptotic theory valid in the limit of small Rossby number $R_o = U / (L \Omega) \ll 1$, with $U$ a typical velocity and $L$ a typical length scale. 
The theory, developed for  three-wave interactions, predicts anisotropic turbulence with a direct cascade preferentially in the direction transverse to the axis of rotation $\boldsymbol{\Omega}$. In the limit $k_\perp \gg k_\|$, where $k_\perp$ and $k_\|$ refer to the wavenumbers perpendicular and parallel to $\boldsymbol{\Omega}$ respectively, an exact solution is derived -- the KZ energy spectrum -- which is $E_k \sim k_\perp^{-5/2} \left\vert k_\| \right\vert^{-1/2}$. This solution corresponds to a stationary state for which the energy flux is constant and positive. Note that there is another type of solution, the Rayleigh-Jeans spectrum associated with the thermodynamical equilibrium of the system and for which the energy flux is zero. 

Rotating hydrodynamic turbulence has been extensively studied both numerically \citep{Bellet_2006, Pouquet_2010, Pouquet_2013, LeReun2017, Buzzicotti2018, Seshasayanan_2018, Sharma_2019, van-Kan_2020} and experimentally \citep{Baroud2002, Morize_2005, vanBokhoven2009, Lamriben_2011, Yarom_2014, Campagne_2014, Godeferd2015, Yarom_2017}. In particular, recent studies \citep{LeReun2020,Monsalve_2020,Yokoyama2021} have reported energy spectra consistent with the prediction of inertial wave turbulence theory.  
But until now, an important theoretical point has been left out: the verification of the locality of the KZ spectrum to ensure the finiteness of the energy flux. This is a criterion of locality of interactions that supports Kolmogorov's idea that the inertial range is independent of the largest (forcing) and smallest (dissipation) scales. To prove the locality of the KZ spectrum, it is necessary to return to the kinetic equation and study the convergence of the integrals \citep{Zakharov_1992}; this is the first objective of this article. The second objective is to study the energy flux to find an estimate of the Kolmogorov constant. 

After a brief introduction to inertial wave turbulence in $\S$ \ref{sec:Locality}, we prove that the KZ spectrum is indeed 'local', which gives strong theoretical support to recent numerical and experimental studies. In $\S$ \ref{sec:Kolmogorov}, we show numerically that the energy fluxes in the perpendicular and parallel directions do not behave in the same way because in the latter case some triadic interactions can have a negative contribution to the energy flux, whereas in the former case all interactions contribute to a positive flux. We also numerically estimate the Kolmogorov constant before concluding in  $\S$  \ref{sec:Conclusion}.

\section{Locality conditions}\label{sec:Locality}
\subsection{Kinetic equation and KZ spectrum}
The inviscid equations for incompressible flows in a rotating frame read
\begin{equation}\label{NavierStokes}
    \pdt \bs{w} + \bs{u} \cdot \bnabla \bs{w} 
    = \bs{w} \cdot \bnabla \bs{u} + 2 \bs{\Omega} \cdot \bnabla \bs{u},
\end{equation}
where $\bs{u}$ is a solenoidal velocity ($\div{\bs{u}}=0$), $\bs{w} = \rot{\bs{u}}$ the vorticity and $\bs{\Omega} = \Omega \hat{\bs{e}}_\|$ ($\vert \hat{\bs{e}}_\| \vert=1$) the constant rotation rate. 
The linear solutions of equation (\ref{NavierStokes}) are inertial waves with the angular frequency
\eq{linearPulsation}{
\omega_k = 2 \Omega \frac{k_\|}{k}.}

The first main result of the theory of inertial wave turbulence is the derivation of the kinetic equation which describes the nonlinear evolution of the energy spectrum  on a time scale much larger than the wave period $\tau \sim 1/\omega_k$. 
For reasons of simplicity, we assume an equipartition of energy densities between the co- and counter-propagating inertial waves, and we therefore do not consider the kinetic helicity which is then zero.
This turbulence being anisotropic with a transfer mainly in the transverse direction to $\boldsymbol{\Omega}$, it is relevant to write the kinetic equation in the limit $k_\perp \gg k_\|$.
Note that we often implicitly assume that the energy is initially isotropically distributed at the largest scales of the system but the same dynamics is expected if initially the energy is located elsewhere, as long as the inertial wave turbulence condition on the time scales is satisfied (see discussion in \cite{Galtier_2003a}). The exception is when the energy is confined to $k_\|=0$ which cannot be described by wave turbulence.
In the limit $k_\perp \gg k_\|$, the kinetic equation reads (for more details, see \cite{Galtier_2003a})
\eq{energy}{
\begin{split}
    \pdt E_k &= \frac{\epsilon^2}{32 \Omega} \sum_{s_k s_p s_q} \int_{\Delta_\perp} 
    \sin \theta
    \frac{s_k s_p p_\|}{p_\perp^2 q_\perp^2 k_\|} \left(s_q q_\perp - s_p p_\perp\right)^2 \left( s_k k_\perp + s_p p_\perp + s_q q_\perp \right)^2 E_q \\
    &\times \left( p_\perp E_k - k_\perp E_p \right) \delta \left(\frac{s_k k_\|}{k_\perp} +  \frac{s_p p_\|}{p_\perp} + \frac{s_q q_\|}{q_\perp} \right) \delta \left( k_\| + p_\| + q_\| \right) \dd p_\perp \dd q_\perp \dd p_\| \dd q_\|,
\end{split}
}
where $E_k = E(k_\perp,k_\parallel)$ is the axisymmetric energy spectrum, $s_i$ is the wave polarity ($s_i =\pm 1$ with $i=k,p,q$), $\theta$ is the angle between $\bs{k}_\perp$ and $\bs{p}_\perp$ in the triangle $\bs{k}_\perp + \bs{p}_\perp + \bs{q}_\perp =  \bs{0}$, $\delta$ is the Dirac distribution, and the integration in the perpendicular direction is done over the domain $\Delta_\perp$ which verifies the previous triadic relation. The integral is preceded by a factor $\epsilon^2$, with $\epsilon$ a small dimensionless parameter ($0 < \epsilon \ll 1$), which can be identified as the Rossby number (which also means that $w \ll \Omega$). This means that the nonlinear dynamics develops on a time scale $\sim \tau /\epsilon^2$, which is much longer than the wave period.
The stationary solution of the kinetic equation (\ref{eq:energy}) is the KZ spectrum 
\eq{KZspectrum}{
E_k = C_E k_\perp^{-5/2} \left\vert k_\| \right\vert^{-1/2},
}
where $C_E$ is necessarily positive (it will be defined later).

\subsection{Convergence domain}

\begin{figure}
    \centering
    \begin{tikzpicture}
        \begin{axis}[small,
            ymin=-1,ymax=2,xmin=0,xmax=5,
            xlabel=$x$, ylabel=$y$, ylabel style={rotate=-90}
            ]
            \addplot[name path global=firstline,  domain=0:4.01]{2-x/2};
            \addplot[name path global=secondline, domain=4:5]{4-x};
            \addplot[name path global=thirdline,  domain=1:5]{3/2-x/2};
            \addplot[name path global=fourthline, domain=0:1.001]{2-x};
            \fill[name intersections={of=firstline and secondline,by=point1},
            name intersections={of=secondline and thirdline,by=point2},
            name intersections={of=thirdline and fourthline,by=point3},
            name intersections={of=fourthline and firstline,by=point4},
            ][fill={rgb, 255:red, 0; green, 64; blue, 221}  ,fill opacity=0.25 ](point1)--(point2)--(point3)--(point4)--(point1);
            \node[circle,fill,inner sep=2pt][fill={rgb, 255:red, 0; green, 64; blue, 221}  ,fill opacity=1 ] at (axis cs:2.5,0.5) {};
        \end{axis}
    \end{tikzpicture}
    \caption{Domain of convergence (where the energy flux is finite) of the kinetic equation for power law solutions $E_k \sim k_\perp^{-x} \left\vert k_\| \right\vert^{-y}$. The blue disk at the center of the domain corresponds to the KZ spectrum.}
    \label{fig:convergenceDomain}
\end{figure}
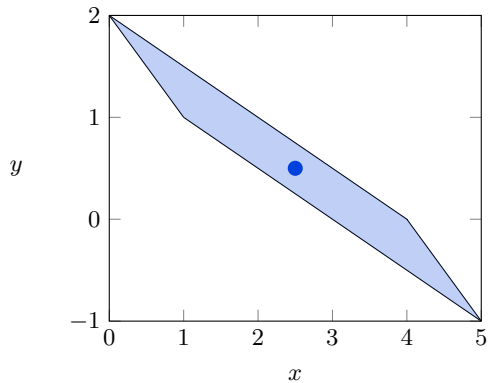

{\bf Proposition:} The domain of convergence of the kinetic equation (\ref{eq:energy}) for power law spectra $E_k \sim k_\perp^{-x} \left\vert k_\| \right\vert^{-y}$ is given by the following locality conditions
\begin{eqnarray}
    3 < x &+& 2 y < 4, \\
2 < x &+& y < 4 ,
\end{eqnarray}
both of which must be satisfied.
This result shows a familiar property of wave turbulence: the KZ spectrum, for which $x=5/2$ and $y=1/2$, falls exactly in the middle of the convergence domain (see Figure \ref{fig:convergenceDomain} for an illustration). 

\begin{figure}
    \centering
\tikzset{every picture/.style={line width=0.75pt}} 

\begin{tikzpicture}[x=0.74pt,y=0.74pt,yscale=-0.66,xscale=0.66]

\draw    (25.13,14.75) -- (25.13,259) ;
\draw [shift={(25.13,11.75)}, rotate = 90] [fill={rgb, 255:red, 0; green, 0; blue, 0 }  ][line width=0.08]  [draw opacity=0] (8.93,-4.29) -- (0,0) -- (8.93,4.29) -- cycle    ;
\draw    (272.13,259) -- (25.13,259) ;
\draw [shift={(275.13,259)}, rotate = 180] [fill={rgb, 255:red, 0; green, 0; blue, 0 }  ][line width=0.08]  [draw opacity=0] (8.93,-4.29) -- (0,0) -- (8.93,4.29) -- cycle    ;
\draw  [color={rgb, 255:red, 0; green, 0; blue, 0 }  ,draw opacity=1 ] (251.25,109.87) -- (103.2,257.93) -- (25.83,180.55) -- (173.88,32.5) -- cycle ;
\draw [color={rgb, 255:red, 255; green, 255; blue, 255 }  ,draw opacity=1 ][line width=1.5]    (173.88,32.5) -- (251.25,109.87) ;
\draw [shift={(251.25,109.87)}, rotate = 45] [color={rgb, 255:red, 255; green, 255; blue, 255 }  ,draw opacity=1 ][fill={rgb, 255:red, 255; green, 255; blue, 255 }  ,fill opacity=1 ][line width=1.5]      (0, 0) circle [x radius= 4.36, y radius= 4.36]   ;
\draw [shift={(173.88,32.5)}, rotate = 45] [color={rgb, 255:red, 255; green, 255; blue, 255 }  ,draw opacity=1 ][fill={rgb, 255:red, 255; green, 255; blue, 255 }  ,fill opacity=1 ][line width=1.5]      (0, 0) circle [x radius= 4.36, y radius= 4.36]   ;
\draw    (343,250) -- (397,250) ;
\draw [shift={(400,250)}, rotate = 180] [fill={rgb, 255:red, 0; green, 0; blue, 0 }  ][line width=0.08]  [draw opacity=0] (8.93,-4.29) -- (0,0) -- (8.93,4.29) -- cycle    ;
\draw    (400,250) -- (371.91,45.97) ;
\draw [shift={(371.5,43)}, rotate = 82.16] [fill={rgb, 255:red, 0; green, 0; blue, 0 }  ][line width=0.08]  [draw opacity=0] (8.93,-4.29) -- (0,0) -- (8.93,4.29) -- cycle    ;
\draw    (371.5,43) -- (343.41,247.03) ;
\draw [shift={(343,250)}, rotate = 277.84] [fill={rgb, 255:red, 0; green, 0; blue, 0 }  ][line width=0.08]  [draw opacity=0] (8.93,-4.29) -- (0,0) -- (8.93,4.29) -- cycle    ;

\draw    (473,250) -- (527,250) ;
\draw [shift={(530,250)}, rotate = 180] [fill={rgb, 255:red, 0; green, 0; blue, 0 }  ][line width=0.08]  [draw opacity=0] (8.93,-4.29) -- (0,0) -- (8.93,4.29) -- cycle    ;
\draw    (530,250) -- (501.91,45.97) ;
\draw [shift={(501.5,43)}, rotate = 82.16] [fill={rgb, 255:red, 0; green, 0; blue, 0 }  ][line width=0.08]  [draw opacity=0] (8.93,-4.29) -- (0,0) -- (8.93,4.29) -- cycle    ;
\draw    (501.5,43) -- (473.41,247.03) ;
\draw [shift={(473,250)}, rotate = 277.84] [fill={rgb, 255:red, 0; green, 0; blue, 0 }  ][line width=0.08]  [draw opacity=0] (8.93,-4.29) -- (0,0) -- (8.93,4.29) -- cycle    ;

\draw    (603,250) -- (657,250) ;
\draw [shift={(660,250)}, rotate = 180] [fill={rgb, 255:red, 0; green, 0; blue, 0 }  ][line width=0.08]  [draw opacity=0] (8.93,-4.29) -- (0,0) -- (8.93,4.29) -- cycle    ;
\draw    (660,250) -- (631.91,45.97) ;
\draw [shift={(631.5,43)}, rotate = 82.16] [fill={rgb, 255:red, 0; green, 0; blue, 0 }  ][line width=0.08]  [draw opacity=0] (8.93,-4.29) -- (0,0) -- (8.93,4.29) -- cycle    ;
\draw    (631.5,43) -- (603.41,247.03) ;
\draw [shift={(603,250)}, rotate = 277.84] [fill={rgb, 255:red, 0; green, 0; blue, 0 }  ][line width=0.08]  [draw opacity=0] (8.93,-4.29) -- (0,0) -- (8.93,4.29) -- cycle    ;

\draw [color={rgb, 255:red, 255; green, 255; blue, 255 }  ,draw opacity=1 ][line width=1.5]  [dash pattern={on 5.63pt off 4.5pt}]  (218.33,142.83) -- (251.25,109.87) ;
\draw [color={rgb, 255:red, 0; green, 64; blue, 221 }  ,draw opacity=1 ][fill={rgb, 255:red, 10; green, 132; blue, 255 }  ,fill opacity=0 ][line width=0.75]    (115.8,90.6) -- (193.18,167.98) ;
\draw [shift={(193.18,167.98)}, rotate = 45] [color={rgb, 255:red, 0; green, 64; blue, 221 }  ,draw opacity=1 ][fill={rgb, 255:red, 0; green, 64; blue, 221 }  ,fill opacity=1 ][line width=0.75]      (0, 0) circle [x radius= 3.35, y radius= 3.35]   ;
\draw [shift={(115.8,90.6)}, rotate = 45] [color={rgb, 255:red, 0; green, 64; blue, 221 }  ,draw opacity=1 ][fill={rgb, 255:red, 0; green, 64; blue, 221 }  ,fill opacity=1 ][line width=0.75]      (0, 0) circle [x radius= 3.35, y radius= 3.35]   ;
\draw  [color={rgb, 255:red, 0; green, 0; blue, 0 }  ,draw opacity=0 ][fill={rgb, 255:red, 0; green, 64; blue, 221 }  ,fill opacity=1 ] (131.54,105.63) -- (145.79,91.39) -- (143.5,89.1) -- (157.57,84.17) -- (152.64,98.24) -- (150.36,95.96) -- (136.11,110.21) -- cycle ;
\draw  [color={rgb, 255:red, 0; green, 0; blue, 0 }  ,draw opacity=0 ][fill={rgb, 255:red, 0; green, 64; blue, 221 }  ,fill opacity=1 ] (151.54,125.63) -- (165.79,111.39) -- (163.5,109.1) -- (177.57,104.17) -- (172.64,118.24) -- (170.36,115.96) -- (156.11,130.21) -- cycle ;
\draw  [color={rgb, 255:red, 0; green, 0; blue, 0 }  ,draw opacity=0 ][fill={rgb, 255:red, 0; green, 64; blue, 221 }  ,fill opacity=1 ] (171.54,145.63) -- (185.79,131.39) -- (183.5,129.1) -- (197.57,124.17) -- (192.64,138.24) -- (190.36,135.96) -- (176.11,150.21) -- cycle ;

\draw  [draw opacity=0][fill={rgb, 255:red, 0; green, 64; blue, 221 }  ,fill opacity=0.25 ] (193.52,167.61) -- (103.2,257.93) -- (25.83,180.55) -- (116.14,90.24) -- cycle ;
\draw [color={rgb, 255:red, 255; green, 255; blue, 255 }  ,draw opacity=1 ][line width=1.5]  [dash pattern={on 5.63pt off 4.5pt}]  (140.96,65.46) -- (173.88,32.5) ;

\draw (246,261.34) node [anchor=north west][inner sep=0.75pt]    {$\tilde{p}_{\perp }$};
\draw (2,13.4) node [anchor=north west][inner sep=0.75pt]    {$\tilde{q}_{\perp }$};
\draw (11.13,174.4) node [anchor=north west][inner sep=0.75pt]    {$1$};
\draw (98.13,261.4) node [anchor=north west][inner sep=0.75pt]    {$1$};
\draw (95,229) node [anchor=north west][inner sep=0.75pt]   [align=left] {A};
\draw (40,172) node [anchor=north west][inner sep=0.75pt]   [align=left] {B};
\draw (188,75) node [anchor=north west][inner sep=0.75pt]   [align=left] {C};
\draw (341,136) node [anchor=north west][inner sep=0.75pt]    {$1$};
\draw (390,133) node [anchor=north west][inner sep=0.75pt]    {$\tilde{p}_{\perp }$};
\draw (360,252) node [anchor=north west][inner sep=0.75pt]    {$\tilde{q}_{\perp }$};
\draw (363,17) node [anchor=north west][inner sep=0.75pt]   [align=left] {A};
\draw (471,136) node [anchor=north west][inner sep=0.75pt]    {$1$};
\draw (520.87,133) node [anchor=north west][inner sep=0.75pt]    {$\tilde{q}_{\perp }$};
\draw (490,252) node [anchor=north west][inner sep=0.75pt]    {$\tilde{p}_{\perp }$};
\draw (493,17) node [anchor=north west][inner sep=0.75pt]   [align=left] {B};
\draw (594,133) node [anchor=north west][inner sep=0.75pt]    {$\tilde{q}_{\perp }$};
\draw (651,133) node [anchor=north west][inner sep=0.75pt]    {$\tilde{p}_{\perp }$};
\draw (620,253) node [anchor=north west][inner sep=0.75pt]    {$1$};
\draw (623,17) node [anchor=north west][inner sep=0.75pt]   [align=left] {C};
\draw (45,133) node [anchor=north west][inner sep=0.75pt]  [rotate=-315]  {$\tilde{p}_{\perp } +1$};
\draw (45.44,204.33) node [anchor=north west][inner sep=0.75pt]  [rotate=-45]  {$1-\tilde{p}_{\perp }$};
\draw (140,225) node [anchor=north west][inner sep=0.75pt]  [rotate=-315]  {$\tilde{p}_{\perp } -1$};
\draw (136.83,115.74) node [anchor=north west][inner sep=0.75pt]  [color={rgb, 255:red, 0; green, 64; blue, 221 }  ,opacity=1 ,rotate=-45]  {$a-\tilde{p}_{\perp }$};
\end{tikzpicture}
    \caption{
    Left: The kinetic equation (\ref{eq:energyIntegree}) is integrated over a domain verifying $\bs{k}_\perp + \bs{p}_\perp + \bs{q}_\perp = \bs{0}$, which corresponds to an infinite band where the boundaries are flattened triangles. Regions A, B and C (at infinity) are those for which the triadic interactions are non-local. 
    We define $\tilde{q}_\perp = a - \tilde{p}_\perp, \: \forall a \in [1,+\infty[$, to restrict the numerical integration to the domain $\Delta_\perp$. Right: Representation of the non-local triadic interactions for regions A, B and C.}
    \label{fig:Algorithm}
\end{figure}
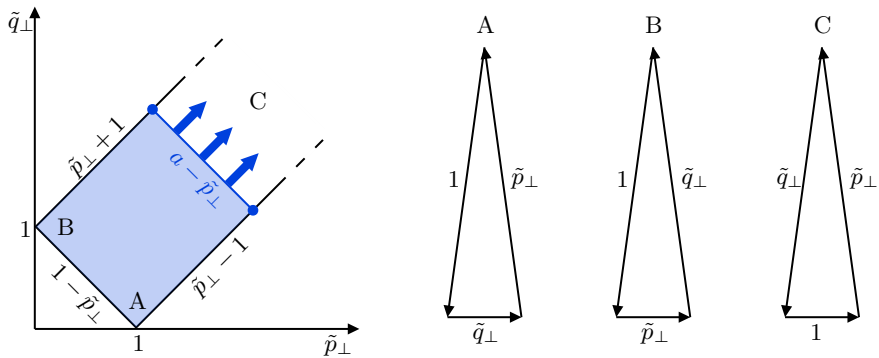

{\bf Proof:}
Starting from equation (\ref{eq:energy}), we seek solutions of the form $E_k = C_E k_\perp^{-x} \left\vert k_\| \right\vert^{-y}$ and introduce the dimensionless variables  $\tilde{p}_\perp = p_\perp k_\perp^{-1}$, $\tilde{q}_\perp = q_\perp k_\perp^{-1}$, $\tilde{p}_\| = p_\| k_\|^{-1} $ and $\tilde{q}_\| = q_\| k_\|^{-1}$. After integration over the parallel wavenumbers, by taking advantage of the following Dirac distribution property
\eq{}{
\int_\mathbb{R} f(x) \delta \left( g(x) \right) \dd x = \sum_i \frac{f \left( x_i \right) }{\left| g' \left( x_i \right) \right|} \quad \text{such as} \quad g \left( x_i \right)=0,
}
to find
\eq{}{
\tilde{p}_\| = \tilde{p}_\perp \frac{s_k \tilde{q}_\perp - s_q}{s_q \tilde{p}_\perp - s_p \tilde{q}_\perp} \quad , \quad \tilde{q}_\| = \tilde{q}_\perp \frac{s_p - s_k \tilde{p}_\perp}{s_q \tilde{p}_\perp - s_p \tilde{q}_\perp},
}
one gets
\eq{energyIntegree}{
\begin{split}
    \pdt E_k =& \frac{\epsilon^2 C_E^2}{32\Omega} k_\perp^{4-2x} \left\vert k_\| \right\vert^{-2y} \sum_{s_k s_p s_q} \int_{\Delta_\perp} s_k s_p  \tilde{q}_\perp^{-x-y-2} \left(s_q\tilde{q}_\perp - s_p \tilde{p}_\perp  \right)^2 \left(s_k + s_p \tilde{p}_\perp + s_q \tilde{q}_\perp  \right)^2 \\
    &\times \sin \theta \frac{s_k \tilde{q}_\perp - s_q}{s_q \tilde{p}_\perp - s_p \tilde{q}_\perp} \left\vert \frac{s_p - s_k \tilde{p}_\perp}{s_q \tilde{p}_\perp - s_p \tilde{q}_\perp} \right\vert^{-y} \left(1 - \tilde{p}_\perp^{-x-y-1} \left\vert \frac{s_k \tilde{q}_\perp - s_q}{s_q \tilde{p}_\perp - s_p \tilde{q}_\perp} \right\vert^{-y} \right) \\
    &\times \left\vert \frac{ \tilde{p}_\perp \tilde{q}_\perp }{s_p \tilde{q}_\perp - s_q \tilde{p}_\perp} \right\vert \dd \tilde{p}_\perp \dd \tilde{q}_\perp,
\end{split}
}
where $\sin \theta= \sqrt{1 - \left(1 + \tilde{p}_\perp^2 - \tilde{q}_\perp^2 \right)^2 \left(2 \tilde{p}_\perp \right)^{-2}}$. 
There are three regions where the triadic interactions are non-local (regions A, B and C in Figure \ref{fig:Algorithm}) and therefore where convergence of the integrals in expression (\ref{eq:energyIntegree}) must be checked.
We now establish the convergence criteria for each of these three regions.

\subsubsection{Region A}

We define $\tilde{p}_\perp = 1 + r \cos \beta$ and $\tilde{q}_\perp = r \sin \beta$, with $r \ll 1$ and $\beta \in \left[\pi/4, 3 \pi/4 \right]$ the polar coordinates with their origin at $\left(\tilde{p}_\perp, \tilde{q}_\perp\right)=(1,0)$. Two cases must be distinguished: $s_k=s_p$ and $s_k=-s_p$.
An evaluation (to leading order) of the different terms of the integral (\ref{eq:energyIntegree}) is given in Table 1. Note that these evaluations take into account the possible cancellation of the integral due to $\beta$ symmetry. 
\begin{center}
\begin{tabular}{ |c|c|c| } 
 \hline
Table 1 & $s_k=s_p$ & $s_k=-s_p$ \\ 
 \hline
$\frac{s_k \tilde{q}_\perp - s_q}{s_q \tilde{p}_\perp - s_p \tilde{q}_\perp}$ & $ -1 $ & $ -1$ \\ 
$ \left\vert \frac{s_p - s_k \tilde{p}_\perp}{s_q \tilde{p}_\perp - s_p \tilde{q}_\perp} \right\vert $ & $ r \vert \cos \beta \vert $ & $ 2 $\\
$\left(s_q\tilde{q}_\perp - s_p \tilde{p}_\perp  \right)^2$ & 1 & 1 \\ 
$\left(s_k + s_p \tilde{p}_\perp + s_q \tilde{q}_\perp  \right)^2$ & 4 & $ r^2$ \\ 
$\sin \theta$ &  $r \sqrt{- \cos 2 \beta}$ & $r \sqrt{- \cos 2 \beta}$ \\
$ 1 - \tilde{p}_\perp^{-x-y-1} \left\vert \frac{s_k \tilde{q}_\perp - s_q}{s_q \tilde{p}_\perp - s_p \tilde{q}_\perp} \right\vert^{-y} $ & $ \propto r^2 \cos^2 \beta$ & $ \propto r^2 \cos^2 \beta$ \\ 
$\left\vert \frac{ \tilde{p}_\perp \tilde{q}_\perp }{s_p \tilde{q}_\perp - s_q \tilde{p}_\perp} \right\vert $ & $r \left\vert \sin \beta \right\vert$ & $r \left\vert \sin \beta \right\vert$ \\ 
$\dd \tilde{p}_\perp \dd \tilde{q}_\perp  $ & $r \dd r \dd \beta $ & $r \dd r \dd \beta $ \\ 
 \hline
\end{tabular}
\end{center}
When $s_k=s_p$, the criterion of convergence of the kinetic equation  (\ref{eq:energyIntegree}) will be given by the following integral 
\eq{}{
    \int_0^{R<1} r^{3-x-2y} \dd r \int_{\pi/4}^{3\pi/4} {\left\vert \cos \beta \right\vert^{2-y}} \sqrt{- \cos 2 \beta} \left( \sin \beta \right)^{-1-x-y} \dd \beta .
}
Therefore, there is convergence if $x+2y < 4$.
When $s_k=-s_p$, we have
\eq{}{
    \int_0^{R<1} r^{5-x-y} \dd r \int_{\pi/4}^{3\pi/4} \cos^2 \beta \sqrt{-\cos 2 \beta} \left( \sin \beta \right)^{-1-x-y}  \dd \beta
}
and the convergence is obtained if $x + y < 6$.

\subsubsection{Region B}

We define $\tilde{p}_\perp = r \cos \beta$ and $\tilde{q}_\perp = 1+ r \sin \beta$, with this time $\beta \in \left[-\pi/4, \pi/4 \right]$ the polar coordinates with their origin at $\left(\tilde{p}_\perp, \tilde{q}_\perp\right)=(0,1)$. We have two cases: $s_k=s_q$ and $s_k=-s_q$. An evaluation (to leading order) of the different terms of the integral (\ref{eq:energyIntegree}) is given in Table 2. Note that these evaluations take into account the possible cancellation of the integral due to $\beta$ symmetry. 
\begin{center}
\begin{tabular}{ |c|c|c| } 
 \hline
Table 2 & $s_k=s_q$ & $s_k=-s_q$ \\ 
 \hline
 $\frac{s_k \tilde{q}_\perp - s_q}{s_q \tilde{p}_\perp - s_p \tilde{q}_\perp}$ & $ s_k s_p r^2 \sin^2 \beta $ & $ - 2 s_k s_p $ \\
 $ \left\vert \frac{s_p - s_k \tilde{p}_\perp}{s_q \tilde{p}_\perp - s_p \tilde{q}_\perp} \right\vert$ & $ 1 $ & $ 1 $ \\
$\left(s_q\tilde{q}_\perp - s_p \tilde{p}_\perp  \right)^2$ & 1 & 1 \\ 
$\left(s_k + s_p \tilde{p}_\perp + s_q \tilde{q}_\perp  \right)^2$ & 4 & $ r^2$ \\ 
$\sin \theta$ &  $\sqrt{1 - \tan^2 \beta}$ & $\sqrt{1 - \tan^2 \beta}$ \\
$ 1 - \tilde{p}_\perp^{-x-y-1} \left\vert \frac{s_k \tilde{q}_\perp - s_q}{s_q \tilde{p}_\perp - s_p \tilde{q}_\perp} \right\vert^{-y} $ & $ 1 - \left( r \cos \beta \right)^{-x-y-1} \left\vert r \sin \beta \right\vert^{-y}$ & $ 1 - 2^{-y} \left( r \cos \beta \right)^{-x-y-1} $ \\ 
$\left\vert \frac{ \tilde{p}_\perp \tilde{q}_\perp }{s_p \tilde{q}_\perp - s_q \tilde{p}_\perp} \right\vert $ & $r \left\vert \cos \beta \right\vert$ & $r \left\vert \cos \beta \right\vert$ \\ 
$\dd \tilde{p}_\perp \dd \tilde{q}_\perp  $ & $r \dd r \dd \beta $ & $r \dd r \dd \beta $ \\ 
 \hline
\end{tabular}
\end{center}
When $s_k=s_q$, the criterion of convergence of the kinetic equation  (\ref{eq:energyIntegree}) will be given by the following integral
\eq{}{
    \int_0^{R<1} r^{3-x-2y} \dd r \int_{-\pi/4}^{+\pi/4} \left( \cos \beta \right)^{-x-y} \left\vert \sin \beta  \right\vert^{2-y} \sqrt{1 - \tan^2 \beta} \dd \beta .
}
Therefore, there is convergence if $x+2y < 4$.
When $s_k=-s_q$, we have
\eq{}{
     \int_0^{R<1} r^{3-x-y} \dd r \int_{-\pi/4}^{+\pi/4} \left( \cos \beta \right)^{-x-y} \sqrt{1 - \tan^2 \beta} \dd \beta
}
and the convergence is obtained if $x + y < 4$.

\subsubsection{Region C}

We define $\tilde{p}_\perp = (\tau_2-\tau_1)/2$ and $\tilde{q}_\perp = (\tau_1+\tau_2)/2$, with $-1 \le \tau_1 \le 1$ and $1 \ll \tau_2$. We have two cases: $s_p=s_q$ and $s_p=-s_q$. An evaluation (to leading order) of the different terms of the integral (\ref{eq:energyIntegree}) is given in Table 3. Note that these evaluations take into account the possible cancellation of the integral due to $\tau_1$ symmetry. 
\begin{center}
\begin{tabular}{ |c|c|c| } 
 \hline
 Table 3 & $s_p=s_q$ & $s_p=-s_q$ \\ 
 \hline
 $\frac{s_k \tilde{q}_\perp - s_q}{s_q \tilde{p}_\perp - s_p \tilde{q}_\perp}$ & $ - s_k s_p / 2 $ & $ - s_k s_p / 2 $ \\
 $ \left\vert \frac{s_p - s_k \tilde{p}_\perp}{s_q \tilde{p}_\perp - s_p \tilde{q}_\perp} \right\vert $ & $ \left\vert \tau_2 \tau_1^{-1} \right\vert / 2$ & $ 1/2 $ \\
 $\left(s_q\tilde{q}_\perp - s_p \tilde{p}_\perp  \right)^2$ & $\tau_1^2$ & $\tau_2^2$ \\ 
$\left(s_k + s_p \tilde{p}_\perp + s_q \tilde{q}_\perp  \right)^2$ & $\tau_2^2$  & $ 1 + \tau_1^2$ \\ 
$\sin \theta$  & $\sqrt{1 - \tau_1^2}$ & $\sqrt{1 - \tau_1^2}$ \\
$ 1 - \tilde{p}_\perp^{-x-y-1} \left\vert \frac{s_k \tilde{q}_\perp - s_q}{s_q \tilde{p}_\perp - s_p \tilde{q}_\perp} \right\vert^{-y} $ & $1 - 2^{x+2y+1} \tau_2^{-x-2y-1}  \left\vert \tau_1 \right\vert^y$ & $ 1 - 2^{x+y-1} \tau_2^{-x-y-1}$ \\
$\left\vert \frac{ \tilde{p}_\perp \tilde{q}_\perp }{s_p \tilde{q}_\perp - s_q \tilde{p}_\perp} \right\vert $ & $ \tau_2^2 \left\vert \tau_1^{-1} \right\vert / 4 $ & $ \tau_2 / 4 $ \\
$\dd \tilde{p}_\perp \dd \tilde{q}_\perp  $ & $ \propto \dd \tau_1 \dd \tau_2 $ & $ \propto \dd \tau_1 \dd \tau_2 $ \\ 
 \hline
\end{tabular}
\end{center}
When $s_p=s_q$, the criterion of convergence of the kinetic equation  (\ref{eq:energyIntegree}) 
will be given by the following integral
\eq{}{
    \int_{-1}^{+1} \sqrt{1-\tau_1^2} \left\vert \tau_1 \right\vert^{y+1} \dd \tau_1 \int_{\tau > 1}^{+\infty} \tau_2^{-x-2y+2} \dd \tau_2 .
}
Therefore, there is convergence if $3 < x + 2y$.
When $s_p=-s_q$, we have
\eq{}{
    \int_{-1}^{+1} \left( 1 + \tau_1^2 \right) \sqrt{1-\tau_1^2} \dd \tau_1 \int_{\tau>1}^{+\infty} \tau_2^{-x-y+1} \dd \tau_2,
}
and the convergence is obtained if $2 < x + y$.

In conclusion, a solution is local if the following two conditions are satisfied
\eq{}{
3 < x + 2 y < 4 \quad \text{and} \quad 2 < x + y < 4.
}

\section{Kolmogorov constant}\label{sec:Kolmogorov}
\begin{figure}
    \centering
    \includegraphics[width=\textwidth]{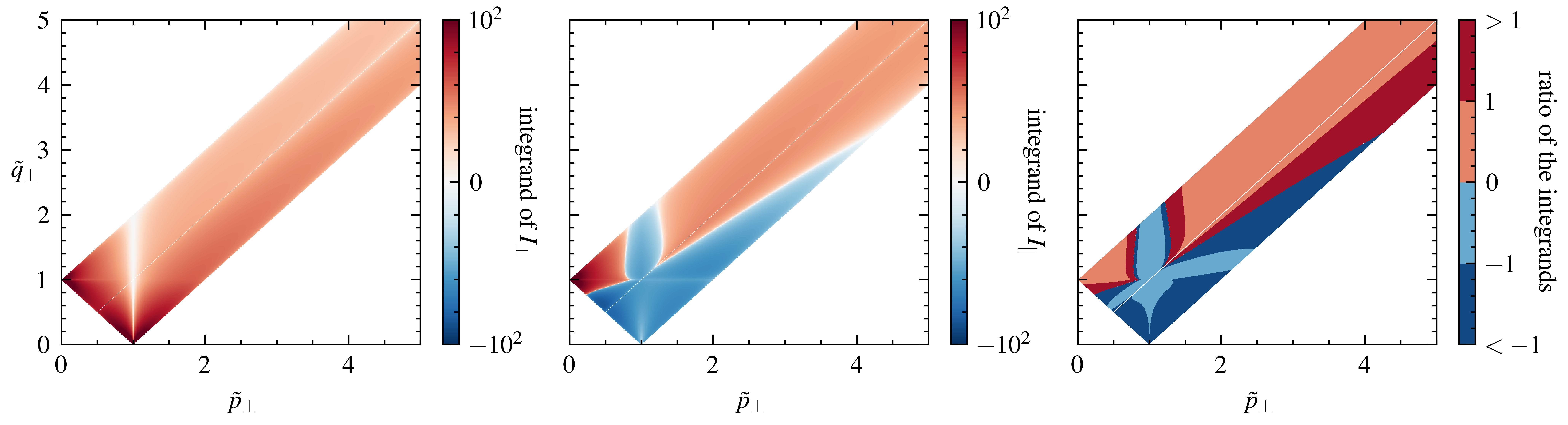}
    \caption{Left: integrand of $I_\perp$ which is always positive. Centre: integrand of $I_\|$ whose sign depends on the perpendicular wavenumbers. Right: integrand of $I_\perp$ divided by the integrand of $I_\|$. The blue colour corresponds to negative values and the red colour to positive values. On the right, the dark colours testify that the integrand of $I_\perp$ is greater than that of $I_\|$ in modulus.}
    \label{fig:flux2D}
\end{figure}

The KZ spectrum is particularly interesting because the associated constant flux allows to measure the so-called Kolmogorov constant $C_K$.
To do so, and following \cite{Zakharov_1992}, we introduce the axisymmetric fluxes
\eq{fluxDefinition}{
\pdt E_k = - \frac{\partial \Pi_\perp (k_\perp,k_\parallel)}{\partial k_\perp} - \frac{\partial \Pi_\|(k_\perp,k_\parallel)}{\partial k_\|} ,
}
and the power law energy spectrum  $E_k = C_E k_\perp^{-x} \left\vert k_\| \right\vert^{-y}$ (we only consider the region $k_\| > 0$). With the dimensionless variables introduced in the previous section, we find after applying the Kuznetsov-Zakharov transformation (see section 3.3 in \cite{Zakharov_1992})
\eq{}{
\pdt E_k = \frac{\epsilon^2 C_E^2}{64 \Omega}  k_\perp^{4-2x} \vert k_\| \vert^{-2y} I(x,y),
}
with
\eq{}{
\begin{split}
    I(x,y) =& \sum_{s_k s_p s_q} \int_{\Delta_\perp} s_k s_p \frac{\tilde{p}_\|}{\tilde{p}_\perp \tilde{q}_\perp^2} \left( s_q \tilde{q}_\perp - s_p \tilde{p}_\perp \right)^2 \left( s_k + s_p \tilde{p}_\perp + s_q \tilde{q}_\perp \right)^2 \sin \theta \\
    &\times  \tilde{q}_\perp^{-x} \left\vert \tilde{q}_\| \right\vert^{-y} \left(1-\tilde{p}_\perp^{-1-x} \left\vert \tilde{p}_\| \right\vert^{-y}\right) \left(1 - \tilde{p}_\perp^{-5+2x} \left\vert \tilde{p}_\| \right\vert^{-1+2y}\right) \\
    &\times \delta \left( s_k + s_p \frac{\tilde{p}_\|}{\tilde{p}_\perp} + s_q \frac{\tilde{q}_\|}{\tilde{q}_\perp} \right) \delta \left( 1 + \tilde{p}_\| + \tilde{q}_\| \right) \dd \tilde{p}_\perp \dd \tilde{q}_\perp \dd \tilde{p}_\| \dd \tilde{q}_\|.
\end{split}
}
After integration, 
and taking the limit $\left( x, y \right) \rightarrow \left( 5/2, 1/2 \right)$, the flux becomes constant and equal to $\Pi_\perp^\mathrm{KZ}$ and $\Pi_\|^\mathrm{KZ}$. Thanks to L'Hospital's rule, we obtain
\eq{fluxKZ}{
    \left( \Pi_\perp^{KZ}  \atop \Pi_\|^{KZ} \right) = \frac{\epsilon^2 C_E^2}{64 \Omega} \left( I_\perp / \vert k_\| \vert \atop I_\| / k_\perp \right),
}
where
\eq{IPerpAndPara}{
\begin{split}
    \left( I_\perp \atop I_\| \right) &\equiv \sum_{s_k s_p s_q} \int_{\Delta_\perp} \frac{s_k s_p \tilde{p}_\|}{\tilde{p}_\perp \tilde{q}_\perp^{9/2}  \left\vert \tilde{q}_\| \right\vert^{1/2}} \left( s_q \tilde{q}_\perp - s_p \tilde{p}_\perp \right)^2 \left( s_k + s_p \tilde{p}_\perp + s_q \tilde{q}_\perp \right)^2 \sin \theta \log \left\vert \left( \tilde{p}_\perp \atop \tilde{p}_\| \right) \right\vert \\
    &\times \left(1- \tilde{p}_\perp^{-7/2} \left\vert \tilde{p}_\| \right\vert^{-1/2}\right) \delta \left( s_k + \frac{s_p \tilde{p}_\|}{\tilde{p}_\perp} +  \frac{s_q \tilde{q}_\|}{\tilde{q}_\perp} \right)  \delta \left( 1 + \tilde{p}_\| + \tilde{q}_\| \right)  \dd \tilde{p}_\perp \dd \tilde{q}_\perp \dd \tilde{p}_\| \dd \tilde{q}_\|.
\end{split}
}
The ratio between the two fluxes $\Pi_\|^\mathrm{KZ} / \Pi_\perp^\mathrm{KZ}$ is proportional to $\vert k_\| \vert / k_\perp$ (which is $\ll 1$) and since $ I_\| / I_\perp \simeq 0.73$ (as checked numerically) then $\Pi_\|^\mathrm{KZ} \ll \Pi_\perp^\mathrm{KZ}$, which is consistent with a turbulent cascade mainly along the perpendicular direction. 
In Figure \ref{fig:flux2D}, we show the sign of the integrands of $I_\perp$ and $I_\|$ obtained from a numerical evaluation of expression (\ref{eq:IPerpAndPara}) after integration over the parallel wavenumbers, and for relatively small perpendicular dimensionless wavenumbers ($<5$).
We see that for $I_\perp$ the integrand is always positive, while for $I_\|$ the integrand can be either positive or negative depending on the perpendicular wavenumbers (for the largest  perpendicular wavenumbers it is always positive), but overall, the positive sign dominates in the sense that $I_\| > 0$.
Therefore, the parallel cascade is also direct but it is composed of different contributions, with some triadic interactions contributing to an inverse cascade.

The energy flux being mostly perpendicular, we consider only the first line of equation (\ref{eq:fluxKZ}) from which we deduce the expression of $C_E$; then, from equation (\ref{eq:KZspectrum}) we obtain (without any additional hypothesis than $k_\perp \gg k_\parallel$)
\eq{definitionCk}{
    E_k = C_K \sqrt{\frac{\Omega \Pi_\perp^\mathrm{KZ}}{\epsilon^2}} k_\perp^{-5/2} \quad \rm{and} \quad C_K = \frac{8}{\sqrt{I_\perp}},
}
with $C_K$ the Kolmogorov constant which can be measured by direct numerical simulation or experimentally. On the theoretical side, to estimate the value of $C_K$ we need to rewrite the expression of $I_\perp$. First of all, we integrate expression (\ref{eq:IPerpAndPara}) over the parallel wavenumbers.
Secondly, because $I_\perp$ is only defined on the domain $\Delta_\perp$, we introduce the change of variable $\tilde{q}_\perp \equiv a-\tilde{p}_\perp$ where $ 1 \le a < +\infty$, and $a-1 \le 2 \tilde{p}_\perp \le a+1$ which confines the integration to this domain (see Figure \ref{fig:Algorithm}). We obtain
\eq{toto}{
\begin{split}
    I_\perp =& \int_{a=1}^{a=+\infty} \int_{\tilde{p}_\perp=(a-1)/2}^{\tilde{p}_\perp=(a+1)/2} \sum_{s_k s_p s_q} \mathcal{H}_{\bs{1} \tilde{\bs{p}} \tilde{\bs{q}}}^{s_k s_p s_q} \dd a \dd \tilde{p}_\perp ,
\end{split}
}
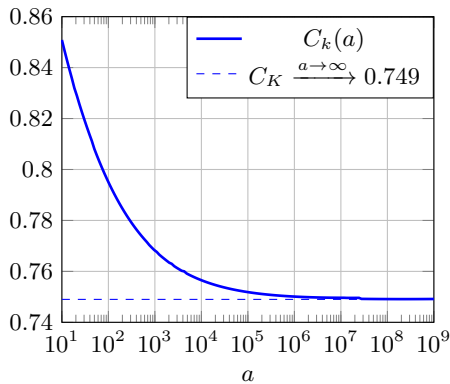
\begin{figure}
    \centering
    \begin{tikzpicture}
\begin{axis}[small,
    legend style={at={(1,1)},anchor=north east},
    ymin=0.74,ymax=0.86,
    xmin=10,xmax=1e9,
	xmode=log, xlabel=$a$,
	grid=major]
\addplot[solid, line width= 1, color=blue] %
	table{dataPlotCk.txt};
\addlegendentry{$C_k (a)$};
\addplot[dashed, color=blue] %
	table{dataPlotC0.txt};
\addlegendentry{$C_K \xrightarrow[]{a \to \infty} 0.749$};
\end{axis}
\end{tikzpicture}
    \caption{Convergence of the numerical estimate of $C_K$ as a function of $a$.}
    \label{fig:ConvergenceCk}
\end{figure}
where
\eq{}{
\begin{split}
    \mathcal{H}_{\bs{1} \tilde{\bs{p}} \tilde{\bs{q}}}^{s_k s_p s_q} &= \frac{s_k s_p \left\vert \frac{s_k (a - \tilde{p}_\perp)-s_q}{\tilde{p}_\perp (s_p+s_q) - s_p a} \right\vert }
    {(a-\tilde{p}_\perp)^{9/2} \sqrt{\left\vert \frac{(a-\tilde{p}_\perp) (s_k \tilde{p}_\perp
    - s_p)}{\tilde{p}_\perp (s_p+s_q)- s_p a }\right\vert}} \left( s_q \left(a - \tilde{p}_\perp \right) - s_p \tilde{p}_\perp \right)^2 \left(s_k + s_p \tilde{p}_\perp + s_q \left(a - \tilde{p}_\perp \right) \right)^2 \\
    &\times \sqrt{1-\frac{\left(a^2-2 a \tilde{p}_\perp-1\right)^2}{4 \tilde{p}_\perp^2}} \left[ 1- \tilde{p}_\perp^{-4} \left(\left\vert \frac{s_q - s_k (a - \tilde{p}_\perp)}{\tilde{p}_\perp (s_p+s_q) - s_p a}\right\vert \right)^{-1/2} \right] \frac{\log \tilde{p}_\perp}{\left\vert \frac{s_p}{\tilde{p}_\perp}-\frac{s_q}{a-\tilde{p}_\perp}\right\vert},
\end{split}
}
is a coefficient with the following symmetries
\begin{equation}
  \mathcal{H}_{\bs{1} \tilde{\bs{p}} \tilde{\bs{q}}}^{s_ks_ps_q} = \mathcal{H}_{\bs{1} \tilde{\bs{p}} \tilde{\bs{q}}}^{-s_k-s_p-s_q},
\end{equation}
which reduces by two the number of integrals to compute (four instead of eight). A numerical integration of (\ref{eq:toto}) gives finally
\begin{equation}
C_K \simeq 0.749 . 
\end{equation}
In Figure \ref{fig:ConvergenceCk}, we show the convergence of $C_K$ to this value as $a$ goes to infinity.

\section{Conclusion and discussion}\label{sec:Conclusion}

In this article, we first derive the locality conditions for inertial wave turbulence and verify that the anisotropic Kolmogorov-Zakharov spectrum -- which is an exact stationary solution of the problem -- satisfies these conditions. The locality of the solution ensures: (i) the finiteness of the associated flux; (ii) the existence of an inertial range independent of large and small scales where forcing and dissipation can occur. Next, neglecting the parallel energy flux, we provide an estimate of the Kolmogorov constant that could be verified by direct numerical simulation or experimentally. Our study completes the one carried out by \cite{Galtier_2003a} and provides theoretical support for recent numerical and experimental studies where energy spectra consistent with the prediction of inertial wave turbulence theory have been reported \citep{LeReun2020,Monsalve_2020,Yokoyama2021}. 

Although the parallel energy flux is positive (like the perpendicular flux), we show numerically that it is in fact the sum of positive and negative contributions. In other words, in the parallel direction, we have triadic interactions that lead to both direct and inverse transfers, with an overall direct cascade. This observation gives some support to experimental results \citep{Morize_2005} showing that the dynamics along the axis of rotation is more difficult to identify than the transverse one, as they exhibit inverse cascade signatures. 
The presence of an inverse cascade in rotating turbulence is often interpreted as the interaction between the slow mode (i.e. the fluctuations at $k_\parallel = 0$) and inertial waves  \citep{Smith_1999}. However, in our case the dynamics of the slow mode is not described by the kinetic equation and the observation made is based solely on the dynamics of inertial wave turbulence. 

The theory of inertial wave turbulence was originally developed for energy and helicity for which two coupled kinetic equations were derived \citep{Galtier_2003a}. Here, helicity is neglected for several reasons. First, it simplifies the mathematical analysis and the physical interpretation. Secondly, in the presence of helicity we cannot derive the Kolmogorov constant because it involves two terms, even if we neglect the parallel flux. Thirdly, numerical simulations are often performed initially without helicity \citep{Bellet_2006}. Some helicity may of course be produced, but it does not usually change the overall dynamics very much. Fourthly, the study with helicity shows that only the state of maximal helicity can give a KZ spectrum, but this situation is not natural as the system quickly breaks this initial assumption, unless it is maintained artificially \citep{Biferale_2012}.

The locality of the KZ spectrum gives some support to the limit of super-local interactions sometimes taken to simplify the study of wave turbulence. Under this limit, a nonlinear diffusion equation is found analytically with which the numerical study becomes much easier \citep{Galtier2020}. Interestingly, it was shown with this model that the non-stationary solution exhibits a $\sim k_\perp^{-8/3}$ spectrum, which is understood as a self-similar solution of the second kind, before forming -- as expected --  the KZ stationary spectrum after a bounce at small scales. 
Note that this diffusion equation is similar to the one derived in plasma physics to study solar wind turbulence \citep{David2019}, which opens the door to laboratory analysis of space plasmas. 

\section*{Declaration of interests} 
The authors report no conflict of interest.

\bibliographystyle{jfm}
\bibliography{main}

\begin{thebibliography}{27}
\expandafter\ifx\csname natexlab\endcsname\relax\def\natexlab#1{#1}\fi
\def\au#1{#1} \def\ed#1{#1} \def\yr#1{#1}\def\at#1{#1}\def\jt#1{\textit{#1}}
  \def\bt#1{#1}\def\bvol#1{\textbf{#1}} \def\vol#1{#1} \def\pg#1{#1}
  \def\publ#1{#1}\def\arxiv#1{#1}\def\org#1{#1}\def\st#1{\textit{#1}}

\bibitem[{Baroud} {\em et~al.\/}(2002){Baroud}, {Plapp}, {She} \&
  {Swinney}]{Baroud2002}
{\sc \au{{Baroud}, C.N.}, \au{{Plapp}, B.B.}, \au{{She}, Z.S.} \&
  \au{{Swinney}, H.L.}} \yr{2002}  \at{{Anomalous Self-Similarity in a
  Turbulent Rapidly Rotating Fluid}}.  \jt{Phys. Rev. Lett.}  \bvol{88}~(11),
  \pg{114501}.

\bibitem[Bellet {\em et~al.\/}(2006)Bellet, Godeferd, Scott \&
  Cambon]{Bellet_2006}
{\sc \au{Bellet, F.}, \au{Godeferd, F.~S.}, \au{Scott, J.~F.} \& \au{Cambon,
  C.}} \yr{2006}  \at{Wave turbulence in rapidly rotating flows}.  \jt{J. Fluid
  Mech.}  \bvol{562},  \pg{83–121}.

\bibitem[{Biferale} {\em et~al.\/}(2012){Biferale}, {Musacchio} \&
  {Toschi}]{Biferale_2012}
{\sc \au{{Biferale}, L.}, \au{{Musacchio}, S.} \& \au{{Toschi}, F.}} \yr{2012}
  \at{{Inverse Energy Cascade in Three-Dimensional Isotropic Turbulence}}.
  \jt{Phys. Rev. Lett.}  \bvol{108}~(16),  \pg{164501}.

\bibitem[{Buzzicotti} {\em et~al.\/}(2018){Buzzicotti}, {Aluie}, {Biferale} \&
  {Linkmann}]{Buzzicotti2018}
{\sc \au{{Buzzicotti}, M.}, \au{{Aluie}, H.}, \au{{Biferale}, L.} \&
  \au{{Linkmann}, M.}} \yr{2018}  \at{{Energy transfer in turbulence under
  rotation}}.  \jt{Phys. Rev. Fluids}  \bvol{3}~(3),  \pg{034802}.

\bibitem[Campagne {\em et~al.\/}(2014)Campagne, Gallet, Moisy \&
  Cortet]{Campagne_2014}
{\sc \au{Campagne, A.}, \au{Gallet, B.}, \au{Moisy, F.} \& \au{Cortet, P.-P.}}
  \yr{2014}  \at{Direct and inverse energy cascades in a forced rotating
  turbulence experiment}.  \jt{Phys. Fluids}  \bvol{26}~(12),  \pg{125112}.

\bibitem[{David} \& {Galtier}(2019)]{David2019}
{\sc \au{{David}, V.} \& \au{{Galtier}, S.}} \yr{2019}  \at{{$k_{\perp}^{-8/3}$
  Spectrum in Kinetic Alfv{\'e}n Wave Turbulence: Implications for the Solar
  Wind}}.  \jt{Astrophys. J. Lett.}  \bvol{880}~(1),  \pg{L10}.

\bibitem[Galtier(2003)]{Galtier_2003a}
{\sc \au{Galtier, S.}} \yr{2003}  \at{Weak inertial-wave turbulence theory}.
  \jt{Phys. Rev. E}  \bvol{68},  \pg{015301}.

\bibitem[Galtier(2023)]{Galtier_2023}
{\sc \au{Galtier, S.}} \yr{2023} {\em Physics of wave turbulence\/}.
  \publ{Cambridge University Press}.

\bibitem[{Galtier} \& {David}(2020)]{Galtier2020}
{\sc \au{{Galtier}, S.} \& \au{{David}, V.}} \yr{2020}
  \at{{Inertial/kinetic-Alfv{\'e}n wave turbulence: A twin problem in the limit
  of local interactions}}.  \jt{Phys. Rev. Fluids}  \bvol{5}~(4),  \pg{044603}.

\bibitem[{Godeferd} \& {Moisy}(2015)]{Godeferd2015}
{\sc \au{{Godeferd}, F.S.} \& \au{{Moisy}, F.}} \yr{2015}  \at{{Structure and
  Dynamics of Rotating Turbulence: A Review of Recent Experimental and
  Numerical Results}}.  \jt{Appl. Mech. Rev.}  \bvol{67}~(3),  \pg{030802}.

\bibitem[van Kan \& Alexakis(2020)]{van-Kan_2020}
{\sc \au{van Kan, A.} \& \au{Alexakis, A.}} \yr{2020}  \at{Critical transition
  in fast-rotating turbulence within highly elongated domains}.  \jt{J. Fluid
  Mech.}  \bvol{899},  \pg{A33}.

\bibitem[{Lamriben} {\em et~al.\/}(2011){Lamriben}, {Cortet} \&
  {Moisy}]{Lamriben_2011}
{\sc \au{{Lamriben}, C.}, \au{{Cortet}, P.-P.} \& \au{{Moisy}, F.}} \yr{2011}
  \at{{Direct Measurements of Anisotropic Energy Transfers in a Rotating
  Turbulence Experiment}}.  \jt{Phys. Rev. Lett.}  \bvol{107}~(2),
  \pg{024503}.

\bibitem[{Le Reun} {\em et~al.\/}(2017){Le Reun}, {Favier}, {Barker} \& {Le
  Bars}]{LeReun2017}
{\sc \au{{Le Reun}, T.}, \au{{Favier}, B.}, \au{{Barker}, A.J.} \& \au{{Le
  Bars}, M.}} \yr{2017}  \at{{Inertial Wave Turbulence Driven by Elliptical
  Instability}}.  \jt{Phys. Rev. Lett.}  \bvol{119}~(3),  \pg{034502}.

\bibitem[{Le Reun} {\em et~al.\/}(2020){Le Reun}, {Favier} \& {Le
  Bars}]{LeReun2020}
{\sc \au{{Le Reun}, T.}, \au{{Favier}, B.} \& \au{{Le Bars}, M.}} \yr{2020}
  \at{{Evidence of the Zakharov-Kolmogorov spectrum in numerical simulations of
  inertial wave turbulence}}.  \jt{EPL (Europhysics Letters)}  \bvol{132}~(6),
  \pg{64002}.

\bibitem[Monsalve {\em et~al.\/}(2020)Monsalve, Brunet, Gallet \&
  Cortet]{Monsalve_2020}
{\sc \au{Monsalve, E.}, \au{Brunet, M.}, \au{Gallet, B.} \& \au{Cortet, P.-P.}}
  \yr{2020}  \at{Quantitative experimental observation of weak inertial-wave
  turbulence}.  \jt{Phys. Rev. Lett.}  \bvol{125},  \pg{254502}.

\bibitem[Morize {\em et~al.\/}(2005)Morize, Moisy \& Rabaud]{Morize_2005}
{\sc \au{Morize, C.}, \au{Moisy, F.} \& \au{Rabaud, M.}} \yr{2005}
  \at{Decaying grid-generated turbulence in a rotating tank}.  \jt{Phys.
  Fluids}  \bvol{17}~(9),  \pg{095105}.

\bibitem[{Nazarenko}(2011)]{Nazarenko_2011}
{\sc \au{{Nazarenko}, S.}} \yr{2011} {\em Wave Turbulence\/},  \st{Lecture
  Notes in Physics},  \vol{vol. 825}.  \publ{Berlin Springer Verlag}.

\bibitem[Pouquet \& Mininni(2010)]{Pouquet_2010}
{\sc \au{Pouquet, A.} \& \au{Mininni, P.~D.}} \yr{2010}  \at{The interplay
  between helicity and rotation in turbulence: implications for scaling laws
  and small-scale dynamics}.  \jt{Phil. Trans. Royal Soc. A}
  \bvol{368}~(1916),  \pg{1635--1662}.

\bibitem[Pouquet {\em et~al.\/}(2013)Pouquet, Sen, Rosenberg, Mininni \&
  Baerenzung]{Pouquet_2013}
{\sc \au{Pouquet, A}, \au{Sen, A}, \au{Rosenberg, D}, \au{Mininni, P~D} \&
  \au{Baerenzung, J}} \yr{2013}  \at{Inverse cascades in turbulence and the
  case of rotating flows}.  \jt{Physica Scripta}  \bvol{T155},  \pg{014032}.

\bibitem[{Seshasayanan} \& {Alexakis}(2018)]{Seshasayanan_2018}
{\sc \au{{Seshasayanan}, K.} \& \au{{Alexakis}, A.}} \yr{2018}
  \at{{Condensates in rotating turbulent flows}}.  \jt{J. Fluid Mech.}
  \bvol{841},  \pg{434--462}.

\bibitem[Sharma {\em et~al.\/}(2019)Sharma, Verma \& Chakraborty]{Sharma_2019}
{\sc \au{Sharma, M.K.}, \au{Verma, M.K.} \& \au{Chakraborty, S.}} \yr{2019}
  \at{Anisotropic energy transfers in rapidly rotating turbulence}.  \jt{Phys.
  Fluids}  \bvol{31}~(8),  \pg{085117}.

\bibitem[Smith \& Waleffe(1999)]{Smith_1999}
{\sc \au{Smith, L.M.} \& \au{Waleffe, F.}} \yr{1999}  \at{Transfer of energy to
  two-dimensional large scales in forced, rotating three-dimensional
  turbulence}.  \jt{Phys. Fluids}  \bvol{11}~(6),  \pg{1608--1622}.

\bibitem[{van Bokhoven} {\em et~al.\/}(2009){van Bokhoven}, {Clercx}, {van
  Heijst} \& {Trieling}]{vanBokhoven2009}
{\sc \au{{van Bokhoven}, L.J.A.}, \au{{Clercx}, H.J.H.}, \au{{van Heijst},
  G.J.F.} \& \au{{Trieling}, R.R.}} \yr{2009}  \at{{Experiments on rapidly
  rotating turbulent flows}}.  \jt{Phys. Fluids}  \bvol{21}~(9),  \pg{096601}.

\bibitem[{Yarom} {\em et~al.\/}(2017){Yarom}, {Salhov} \& {Sharon}]{Yarom_2017}
{\sc \au{{Yarom}, E.}, \au{{Salhov}, A.} \& \au{{Sharon}, E.}} \yr{2017}
  \at{{Experimental quantification of nonlinear time scales in inertial wave
  rotating turbulence}}.  \jt{Phys. Rev. Fluids}  \bvol{2}~(12),  \pg{122601}.

\bibitem[Yarom \& Sharon(2014)]{Yarom_2014}
{\sc \au{Yarom, E.} \& \au{Sharon, E.}} \yr{2014}  \at{Experimental observation
  of steady inertial wave turbulence in deep rotating flows}.  \jt{Nature
  Physics}  \bvol{10}~(7),  \pg{510--514}.

\bibitem[{Yokoyama} \& {Takaoka}(2021)]{Yokoyama2021}
{\sc \au{{Yokoyama}, N.} \& \au{{Takaoka}, M.}} \yr{2021}  \at{{Energy-flux
  vector in anisotropic turbulence: application to rotating turbulence}}.
  \jt{J. Fluid Mech.}  \bvol{908},  \pg{A17}.

\bibitem[{Zakharov} {\em et~al.\/}(1992){Zakharov}, {L'Vov} \&
  {Falkovich}]{Zakharov_1992}
{\sc \au{{Zakharov}, V.~E.}, \au{{L'Vov}, V.~S.} \& \au{{Falkovich}, G.}}
  \yr{1992} {\em {Kolmogorov spectra of turbulence I: Wave turbulence}\/}.
  \publ{Springer Berlin, Heidelberg}.

\end{thebibliography}

\end{document}